\documentclass[a4paper,aps,pre,twocolumn,groupedaddress,showpacs,floats,floatats,floatfix]{revtex4} 
\usepackage[latin1]{inputenc}
\usepackage{graphicx}  
\usepackage{dcolumn}   
\usepackage{bm}        
\usepackage{natbib}
\usepackage{latexsym}
\usepackage{mathrsfs}
\usepackage{amssymb}
\usepackage{amsmath}
\usepackage{amscd}
\usepackage{color}
\usepackage{verbatim}
\begin{document}
\title{Instability statistics and mixing rates}  
\author{Roberto Artuso${}^{1,2,}$}
\email{roberto.artuso@uninsubria.it} 
\author{Cesar Manchein${}^{1,3,}$}
\email{cmanchein@gmail.com}%
\affiliation{${}^1$Center for Nonlinear and Complex Systems and Dipartimento di Fisica e Matematica,
Universit\`a degli Studi dell'Insubria, Via Valleggio 11, 22100 Como, Italy}

\affiliation{${}^2$I.N.F.N. Sezione di Milano, Via Celoria 16, 20133 Milano, Italy}

\affiliation{${}^3$Departamento de F\'\i sica, Universidade Federal do Paran\'a,
  81531-980 Curitiba, Paran\'a, Brazil} 


\date{\today}

\begin{abstract}
We claim that looking at probability distributions of \emph{finite time} largest 
Lyapunov exponents, and more precisely studying their  large deviation properties, yields 
an extremely powerful technique to get quantitative estimates of polynomial
decay rates of time correlations and Poincar\'e recurrences in the
-quite delicate- case of dynamical systems with weak chaotic
properties. 
\end{abstract}

\pacs{05.45.-a,05.45.Ac}


\maketitle
\section{Introduction}
\label{Introduction}
A general problem of dynamical systems theory concerns both the
evaluation of time or space averages, and an understanding of how
finite order estimates of such averages 
converge to their asymptotic limit. The second feature is 
tightly connected to a quantitative estimation of mixing rates, that
is how time correlations asymptotically decay. 
Though in extremely simplified models (simple Markov chains, lattice
models with a finite transition matrix) the 
connection between convergence of finite order estimates and mixing
properties may be established quite easily, 
the issue in more general contexts is much more complicated, and many
facets of the problem still remain as 
open problems. In particular when weakly chaotic systems are
considered it is expected that sticking to regular 
structures in the phase space severely degrades mixing properties, and
critical, polynomial decay of temporal  
correlations can be observed. We remark that mixed systems are thought
to be generic~\cite{MM}, and that slow polynomial decay may influence
deeply deterministic transport properties~\cite{AC-kl}, as the Kubo
formula suggests. We here propose to look at large deviation
properties of finite time estimates as an efficient tool to get
\emph{quantitative} information about mixing properties of the
system. We emphasize that the idea of a relationship between the tails
of finite time distributions and memory effects is not new: for instance
inspection of the tails of finite time Lyapunov exponents  was used
to get informations about qualitative changes in the dynamics
of coupled maps in~\cite{Vcp}, the detection of small regular islands
in~\cite{steven07,bmr}, or motion of a particle in a random
time-dependent potential \cite{scho} (see also \cite{anteneodo04} for
an example in hyperbolic dynamics). A rigorous analysis, inspiring the
present work,  was presented in ~\cite{luzzatto,Mel}: originally
\cite{luzzatto} a class of one-dimensional maps $f$ was taken into
account (the theoretical analysis was extended to higher dimensional systems in \cite{luzzatto2}: 
if we denote by $\lambda$ the Lyapunov exponent, and by
$P_n(\lambda_n)$ the distribution of finite time estimates 
\begin{equation}
\label{finite-l}
\lambda_n(x_0)\,=\,\frac{1}{n} \ln \left| \left. \frac{d f^{(n)}(x)}{dx}\right|_{x_0} \right|
\end{equation}
then, by fixing a threshold $\tilde{\lambda}$ such that $0<
\tilde{\lambda} < \lambda$ we may estimate the fraction of initial
conditions yielding an estimate below the threshold 
\begin{equation}
\label{fractLD}
{\cal M_{\tilde{\lambda}}}(n)\,=\,\int_{-\infty}^{\tilde{\lambda}}\,d\lambda_n \, P_n(\lambda_n).
\end{equation}
The quantity ${\cal M}_{\tilde{\lambda}}(n)$  asymptotically vanishes
if the system is ergodic (as we always suppose): for weakly chaotic
system it may however decay polynomially 
\begin{equation}
\label{pol-dec-M}
{\cal M}_{\tilde{\lambda}}(n)\,\sim \, \frac{1}{n^{\xi}},
\end{equation}
in such a case a bound is proven~\cite{luzzatto} for correlation decay:
\begin{equation}
\label{ALPbound}
{\cal C}(n) \leq \frac{1}{n^{\xi-1}},
\end{equation}
where, as usual,
\begin{equation}
\nonumber
{\cal C}(n)\,=\, \int d\mu(x) \phi(x) \psi(f^{(n)}(x))
\end{equation}
where $\phi$ and $\psi$ are taken in a suitable class of smooth
observables~\cite{CC-corr} and $\mu$ is the invariant measure of the
system. Actually in \cite{luzzatto} ${\cal M}_{\tilde{\lambda}}(n)$ is  
defined in a slightly different way (a full equivalence is established
only via further assumptions): our definition accomplishes a twofold
purpose: on the one side it allows a comparison with large deviation
estimates in \cite{Mel}, and on the other side it yields an easily
computable quantity.  

This actually leads to the main point of our paper: we argue that
scrutinizing the way in which ${\cal M}_{\tilde{\lambda}}(n)$ decays
provides an extremely efficient way of studying quantitatively the
decay of correlations, a key issue in the analysis of fully and weakly
chaotic dynamical systems. As a matter of fact direct quantitative estimates
of mixing rates are known to be quite delicate
numerically~\cite{artuso99}, and many efforts have been devoted to
devise how to tackle the problem by alternative approaches, like
return time statistics~\cite{CL,CS,Kar,artuso99}, which  can be
rigorously shown to yield the correct answer in 1d
intermittency~\cite{Young}. 

The plan of the paper is as follows: firstly we suggest, in view of
recent results \cite{Mel}, that the estimate (\ref{ALPbound}) is not
optimal, and the decay properties of ${\cal M}_{\tilde{\lambda}}(n)$
and ${\cal C}(n)$ are the same; then we check such a conjecture for a  
class of one-dimensional intermittent maps, where polynomial decay
rates of correlations are known exactly. We then consider an
intermittent area preserving map, which is a prototype 
example of intermittency in higher dimensions. We finally reinvestigate, by this technique, the
problem of \emph{generic} correlation decay in maps with a mixed phase
space: in all cases the results are coherent with exact results, and
corroborate proposed conjectures, in a clean, controlled way. 

\section{Large Deviations and Correlation Function}
\label{ldcf}
The way in which we defined ${\cal M}_{\tilde{\lambda}}(n)$
(\ref{fractLD}) is in the form  of a large deviation
result~\cite{Holl}: a remarkable connection 
between large deviations properties and correlation decay was proven
in \cite{Mel} (see also \cite{PS2}): if we consider a system for 
which correlation of smooth functions decay polynomially (the smoothness 
is essential \cite{CC-corr}, see also \cite{CoIs})
${\cal C}(n) \sim n^{-\xi}$ then the following results hold, for
(C\'esaro) finite time averages of an observable $\phi$: 
\begin{equation}
\mu\left(x \, \bigg| \,|n^{-1}\sum_{k=0}^{n-1} \phi(f^{(k)}(x))-\overline{\phi}| > \epsilon\right) \leq
C_{\phi,\epsilon}\frac{1}{n^{\xi}};
\label{melbLD}
\end{equation}
$\mu$ being  again the invariant measure, while $\overline{\phi}$
denotes the phase average of the observable. 
In one dimension, by taking $\phi(x)=\ln |f'(x)|$ this suggests that
the bound (\ref{ALPbound}) could be sharpened, in such a way that
correlation decay and large deviations are  
characterized by the same power-law exponent: on the other side in
dimensions higher than one finite estimates of Lyapunov exponents
(computed for instance by selecting the largest eigenvalue of the
product of jacobians along a trajectory) lack composition structure 
of C\'esaro sums. Our claim is that, despite these provisos, the
asymptotic decay of ${\cal M}_{\tilde{\lambda}}(n)$ quantitatively
reproduces exactly the decay of correlations, and inspection of such a
quantity provides an excellent tool to investigate the weak-chaos regime. 

Before applying such a method to a number of dynamical systems, we point out that the way
in which we consider the exponent $\xi$ is as the lowest possible value in a suitable class of (zero mean) 
smooth functions: it is well known that specific choices of observables may yield faster decay (see \cite{Gou}
for a rigorous example in one-dimensional intermittent dynamics).

\subsection{One-dimensional system: the Pikovisky map}
\label{pm}
\begin{figure}[htb]
\begin{center}
 \includegraphics*[width=8cm,angle=0]{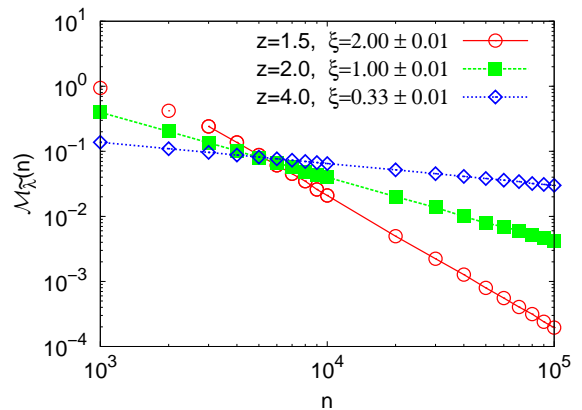}
\end{center}
\caption{Numerical data (symbols), and regression fits (lines) of
  power-law decay of $\mathcal{M}_{\tilde\lambda}(n)$ for Pikovsky map
  for three different values of the intermittency exponent. The fit
  for the case $z=1.5$ starts at $n=3000$. Theoretical values for
  $\xi$, from (\ref{p-corr}), are $\xi=2$, $\xi=1$ and $\xi=1/3$.}  
\label{z1.5}
\end{figure}

Our first benchmark tool is represented by Pikovsky map~\cite{pik91}
$T_z$, which is implicitly defined by 
\begin{eqnarray}
  x=\left\{
\begin{array}{ll}
  \displaystyle\frac{1}{2z}[1+T_z(x)]^z, \hspace{1.6cm} 0 < x < 1/(2z), \\
  T_z(x) + \displaystyle\frac{1}{2z}[1-T_z(x)]^z,  \hspace{0.3cm} 1/(2z) < x < 1;
\end{array}
\right.
\label{pik}
\end{eqnarray}
while for negative values of $x$, the map is defined as
$T_z(-x)=-T_z(x)$~\cite{pik91}. Such a map has remarkable 
features, as in $x= \pm 1$ two marginal fixed points of the
Pomeau-Manneville type are present (and $z$ is 
the corresponding intermittency parameter), while for any value of $z$ 
the invariant measure is Lebesgue~\cite{pik91}, 
as it can be seen by direct inspection of the Perron-Frobenius operator. 
For this map the correlations decay polynomially, with an exponent
that depends upon the intermittency parameter $z$~\cite{CHMV} 
\begin{equation}
  {\cal C}(n) \sim {n^{-1/(z-1)}}.
\label{p-corr}
\end{equation}
This decay is very well reproduced by numerical simulations for ${\cal
  M}_{\tilde{\lambda}}(n)$, as we see from fig. (\ref{z1.5}). We
remark that the numerical simulations for the Pikovsky map were performed 
by using $10^6$ initial conditions uniformly chosen in the phase space $[-1,1]$. 

\begin{figure}[htb]
\begin{center}
\includegraphics*[width=8cm,angle=0]{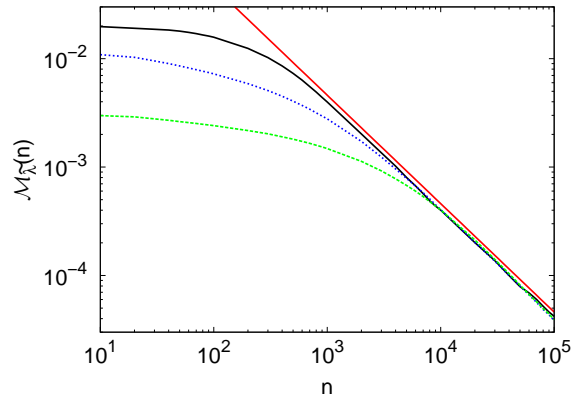}
\end{center}
\caption{Decay of $\mathcal{M}_{\tilde\lambda}(n)$ for the system
  (\ref{pik}) with $z=2.0$, for different cutoffs $\tilde{\lambda}$,
  where the red (gray) curve is a regression fit of power-law decay with
  exponent $\xi=1$.}    
\label{fcut}
\end{figure}

When we want to check how well correlation decay is reproduced by
large deviations, expressed by the decay of ${\cal
  M}_{\tilde{\lambda}}(n)$, we have to be careful to take the 
large deviation parameter $\tilde{\lambda}$ {\it sufficiently far}
from the centre of the distribution: even if the general expectation
is the asymptotic decay law will not depend upon the cutoff, finite
time estimates require that ${\cal M}$ picks up only contributions in
the tail of the distribution. As a matter of fact any choice of the
cutoff $\tilde{\lambda}$ will determine a time scale
$\tau_{\tilde{\lambda}}$ such that only after such a transient the
asymptotic decay is reproduced correctly. This is well illustrated in
fig. (\ref{fcut}) where were chosen three values of 
$\tilde{\lambda}$ and we obtained the asymptotic decay for all cases,
but with different time scale $\tau_{\tilde{\lambda}}$.

\begin{figure}[htb]
\begin{center}
\includegraphics*[width=8cm,angle=0]{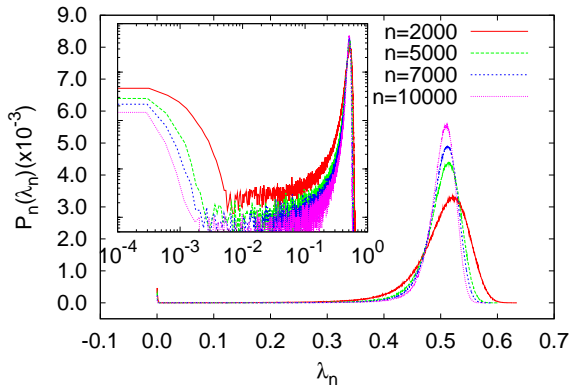}
\end{center}
\caption{Distribution of Lyapunov exponents $P_n(\lambda_n)$
  for the map (\ref{pik}) with $z=2.0$. Inset: a
  magnification of $P_n(\lambda_n)$ is shown in logarithmic scale. The
  distribution of Lyapunov exponents closer to zero show a polynomial
  decay with time.} 
\label{ddist}
\end{figure}

In fig. (\ref{ddist}) we show the shape (bimodal) of (normalized) finite time distribution
$P_n(\lambda_n)$ for the map (\ref{pik}): as ergodicity predicts, they
tend to a Dirac $\delta$ centered on $\lambda$ in the asymptotic
limit, but the tails are polynomially ``fat'' for finite times. 
In the inset (inside fig. (\ref{ddist})) we can detect an interesting feature: by
analyzing the distribution of Lyapunov exponents closer to zero in
logarithmic scale it is possible see how the distribution obeys a power-law decay with
time. 

As a final remark we point out that, though large deviation estimates like (\ref{melbLD}) are 
symmetric with respect to the asymptotic phase average, from a physical point of view it is quite
natural to focus the interest on the small instability branch, as it is precisely anomalous proliferation
of almost stable segments of trajectories that induces transition from exponential to power-law
decay of temporal correlations.

\subsection{Two-dimensional systems: a family of area-preserving maps}
\label{ap}
The second dynamical system we consider is a family of area-preserving
maps on the two-torus $[-\pi, \pi)^2$,  depending upon two parameters
  $\varepsilon$ and $\gamma$: 
\begin{eqnarray}
  M_{(\varepsilon,\gamma)}:\left\{
\begin{array}{ll}
  y_{n+1} = y_n + f(x_n)\qquad & \mathrm{mod} \, 2 \pi ,\\
  x_{n+1} = x_n + y_{n+1}\qquad & \mathrm{mod} \, 2 \pi,
\end{array}
\right.
\label{bimap}
\end{eqnarray}
where $f(x_n)$ is defined by
\begin{equation}
  f(x_n) = [x_n - (1-\varepsilon)\sin(x_n)]^{\gamma}.
\end{equation}

A map of this family was introduced in \cite{lei}, and different
features were analyzed in \cite{ap,liv1,liv2,acc}: we recall a few of
the relevant properties. When $\varepsilon >0$ the map is hyperbolic,
while for $\varepsilon=0$ the fixed point at $(0,0)$ becomes
parabolic: in such a case dynamics close to the fixed point depends
upon the value of $\gamma$, which plays the role of an intermittency
parameter: correspondingly the decay of correlations is exponential in
the former case, while in the latter a power law is expected, with an
exponent depending on $\gamma$. In \cite{acc} the following law was
proposed 
\begin{equation}
{\cal C}(n)\sim n^{-3(\gamma+1)/(3\gamma-1)};
\label{fam-corr}
\end{equation} 
in the case $\gamma=1$ a rigorous bound is proved~\cite{liv2}  $|{\cal
  C}(n)| \leq n^{-2}$ (while (\ref{fam-corr}) predicts an exponent
$-3$).  

\begin{figure}[htb]
\begin{center}
 \includegraphics*[width=8cm,angle=0]{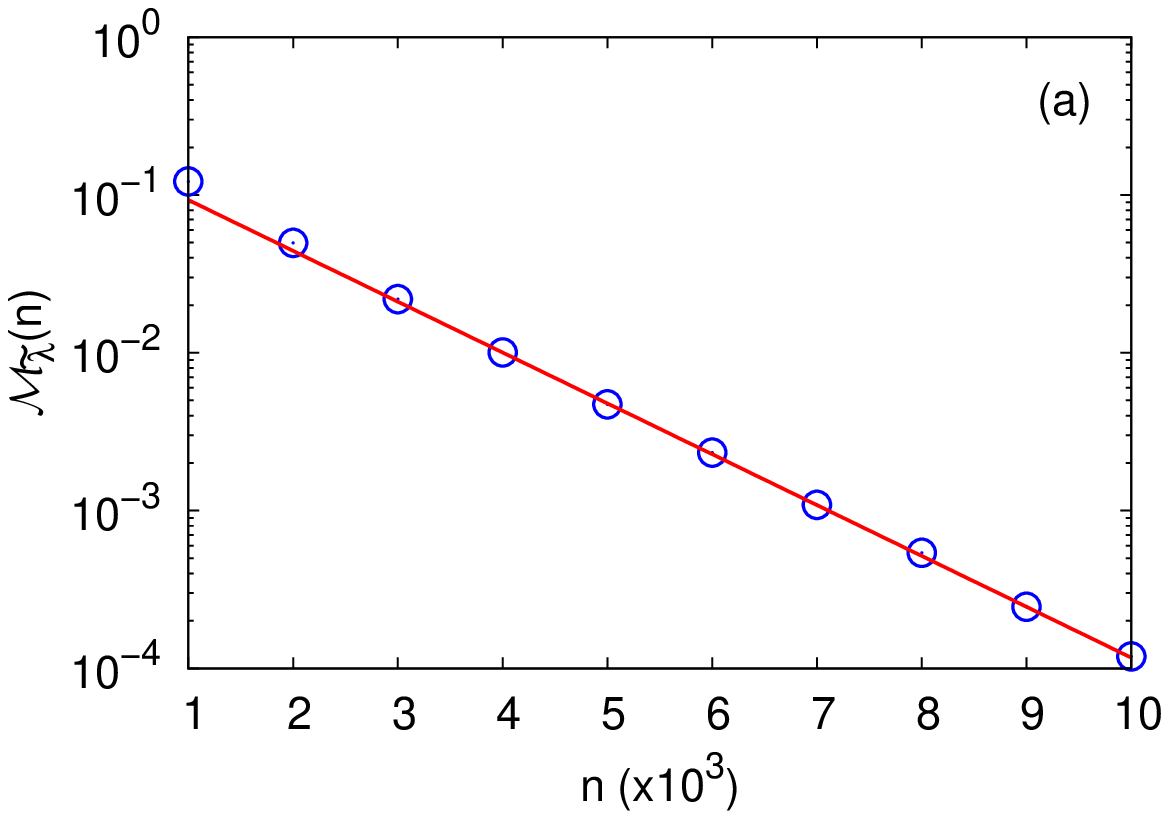}
 \includegraphics*[width=8cm,angle=0]{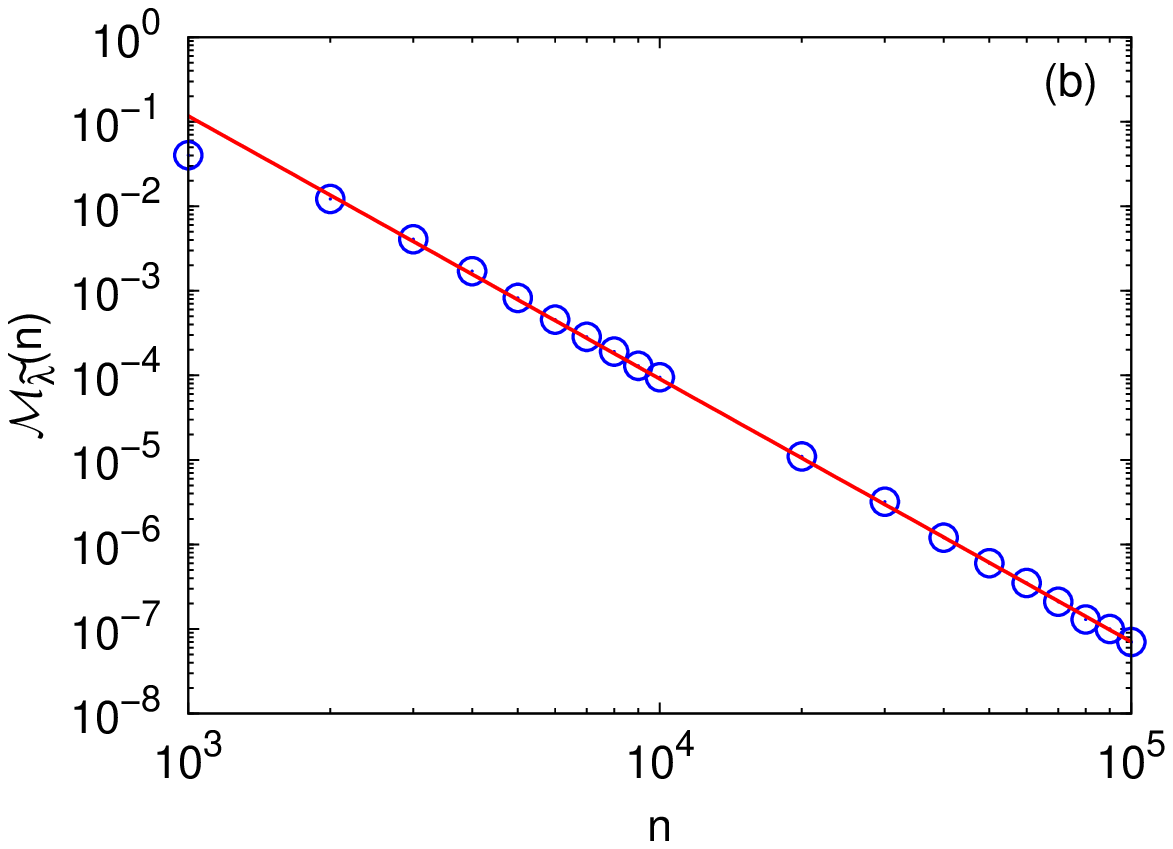}
\end{center}
\caption{Decay $\mathcal{M}_{\tilde{\lambda}}(n)$ (symbols) together
  with a regression fit (full lines) for the map (\ref{bimap}) with
  (a) $\varepsilon = 0.5$ and $\gamma = 1$ (exponential rate decay
  $0.87 \pm 0.01$ ) and (b) $\varepsilon = 0$ and $\gamma = 1$
  (polynomial rate decay $3.05 \pm 0.05$). Both fits were done by
  starting at $n=3000$.} 
\label{thefam}
\end{figure}

Our numerical experiments encompass both regimes: in 
fig.~(\ref{thefam}a) a hyperbolic parameter choice leads to an
exponential decay rate for ${\cal M}_{\tilde{\lambda}}(n)$, while in
the intermittent case a power law is observed, with an exponent in
agreement with (\ref{fam-corr}), see fig. (\ref{thefam}b). 
In this case the numerical results were obtained by using $10^6$
and $10^8$ initial conditions uniformly distributed in the phase
space, respectively.

The case reported in fig.~(\ref{thefam}b) allows to scrutinize the numerical virtues 
of our approach with alternative methods (that were employed in \cite{acc}): it provides
estimates as sharp as the analysis of return time statistics (by employing however much less 
initial conditions), while it outperforms direct computation of correlations.


\subsection{Ensemble of modified standard maps}
\label{sm}
The last case we present concerns a general issue, namely whether
universality properties are exhibited for Hamiltonian systems with a
hierarchical (mixed) phase space. This is a much debated issue (we
refer to \cite{g-r,altmann} for relevant references over the last
twenty years): in particular~\cite{g-r} suggests, on the basis of a
Markov tree model \cite{MO} with random scaling factors for transition
probabilities, the existence of an asymptotic algebraic decay for the
Poincar\'e recurrences. The authors also discuss in detail the
numerical difficulties in probing such a claim: in particular,
following \cite{CSpro,ChGr} they advocate that an average over
different Hamiltonian systems significantly reduces the extremely long
times that are required for a single trajectory to sample fine details
of the phase space structure: this procedure allows them to check for
an ensemble of area preserving maps their proposal, and the polynomial
decay exponent of Poincar\'e recurrences is estimated as $\chi\simeq
1.57 \pm 0.03$, which, according to \cite{CL,CS,Kar,artuso99},
suggests for correlations the decay law 
\begin{equation}
\label{c-gr}
{\cal C}(n) \sim \frac{1}{n^{\chi -1}}.
\end{equation}
The exponent of algebraic decay of Poincar\'e recurrences was also
numerically investigated in \cite{altmann} (with an estimate $\chi
\simeq 1.60 \pm 0.05$), for a different ensemble of area preserving
maps, which we also utilize in the present framework. 

\begin{figure}[htb]
\begin{center}
\includegraphics*[width=8cm,angle=0]{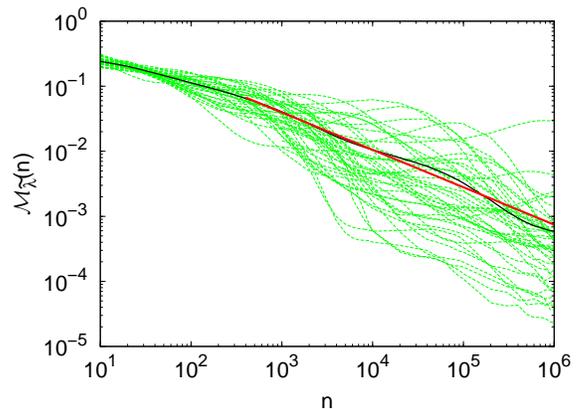}
\end{center}
\caption{Distribution of largest Lyapunov exponents for 40
  realizations of the map (\ref{edmap}). The black curve is the average
  over all curves and the red (gray) line corresponds to the fit of
  average. The fit decay with an exponent $0.57 \pm 0.05$. The fit was
done by starting at $n=500$.} 
\label{hhist}
\end{figure}

The ensemble of modified standard maps on $[-\pi, \pi)^2$ considered
here are given by  
\begin{eqnarray}
  P_{K, K^{\dagger}}:\left\{
\begin{array}{ll}
  p_{n+1} = p_n + K\sin ( x_n) + K^{\dagger}, \\
  x_{n+1} = x_n + p_{n+1}
\end{array}
\right.
\label{edmap}
\end{eqnarray}
where $K$ and $K^{\dagger}$ are the nonlinear parameter and magnetic
field, respectively. They are uniformly chosen in the intervals
$K\in[\pi,1.2 \cdot \pi]$ and  $K^{\dagger}\in [0.0,0.4\cdot \pi]$. We
have used $10^6$ initial conditions taken uniformly in the square $x_0
\in [-0.1,0.1]$, $p_0\in [-0.1,0.1]$, well inside the chaotic region.
The behavior we show in fig.~(\ref{hhist}) apparently arises after a few ($ \sim 10$)
random realizations of the maps: the overall decay rate with an exponent  $0.57 \pm 0.05$
is stable, as well as the oscillations that may be noticed in fig.~(\ref{hhist}). Wether such
oscillations are due to finite size effects, corrections to scaling, or {\em multifractality} \cite{AZ},
is a feature that in our opinion deserves further investigations.
Again we remark that the present method has an
important advantage that it needs a ``small'' number of initial
conditions and realizations of the map to show the asymptotic
decay of $\mathcal{M}_{\tilde\lambda}(n)$ as compared with
traditional procedures using direct computation of correlation functions
or Poincar\'e recurrences.  

\section{Conclusions}
We have shown how the analysis of large deviations for Lyapunov
exponents provides a very efficient way to investigate correlation and
Poincar\'e recurrences decay for dynamical systems: 
in particular when dealing with weakly chaotic systems, with
associated polynomial correlation decay, the power law exponent has
been computed from the distribution function of finite time Lyapunov
exponents with remarkable accuracy. 

We also remark that in all our tests this procedure performs quite
well numerically: for instance in our last example the estimate has
been obtained with a computational effort significantly reduced with
respect to former simulations~\cite{g-r,altmann}. Another interesting
point we plan to investigate in the future is to study the
decay law of correlation function in higher 
dimensional systems where the nonlinearity is not uniformly
distributed along different unstable directions~\cite{mbr}.

\section{Acknowledgments}
One of the authors (CM) gratefully acknowledges CNPq Brazil for a
fellowship, and the Center for Nonlinear and Complex Systems in Como,
for hospitality. We thank Marcus W. Beims, Giulio Casati, Giampaolo
Cristadoro, Stefano Luzzatto, Matteo Sala and Sandro Vaienti for very
useful correspondence and discussions. 

\bibliography{references}

%
\end{document}